# Nonlinear (Anharmonic) Casimir Oscillator

*Habibollah Razmi [a],\*, Mohammad Reza Mohammadi [b], Mahnaz Abdollahi [a], Seyed Mahdi Fazeli [a]*

[a] *Department of Physics, The University of Qom, Qom 37185-359, I.R. Iran*
[b] *Department of Physics, Sistan and Baluchestan University, Zahedan 98135–674, I.R. Iran*

We want to study the dynamics of a simple linear harmonic micro spring which is under the influence of the quantum Casimir force/pressure and thus behaves as a (an) nonlinear (anharmonic) Casimir oscillator. Generally, the equation of motion of this nonlinear micromechanical Casimir oscillator has no exact solvable (analytical) solution and the turning point(s) of the system has (have) no fixed position(s); however, for particular values of the stiffness of the micro spring and at appropriately well-chosen distance scales and conditions, there is (are) approximately sinusoidal solution(s) for the problem (the variable turning points are collected in a very small interval of positions). This, as a simple and elementary plan, may be useful in controlling the Casimir stiction problem in micromechanical devices.

## Introduction

Quantum phenomena have unavoidable roles in the operation of micromechanical systems. One of these phenomena is the Casimir effect [1]. For two parallel plates at a distance of about $10\ nm$ from each other, there is an attractive Casimir force per unit area (Casmir pressure) of about $1\ atm$. The development of powerful experimental methods to measure the Casimir force and its potential importance in micromechanical devices have stimulated new interest in the physical principles underlying the Casimir effect [2-3]. The Casimir force between uncharged metallic surfaces originates from quantum-mechanical zero-point fluctuations of the electromagnetic field. This quantum electro dynamical effect has a profound influence on the oscillatory behavior of microstructures when surfaces are in close proximity. The importance of the Casimir effect in nano systems has been considered since about two decades ago [4]. Because of the strong attractive Casimir force at small scales, moving parts of micromechanical systems may stick to each other; this phenomenon is called stiction. Stiction is a troublemaker effect in micromechanical systems [5-6] and can make them unstable [7]. Here, as a simple and elementary model/plan, we want to know how we can control the stiction problem by connecting oscillating systems (micro springs) between the elements which are under the influence of the Casimir stiction. Clearly, knowing the stiffness coefficient of such oscillating systems is one of the most important necessary information for an engineering management of the problem. We study a simple model of an oscillating system (e.g. a micro spring) under the influence of the Casimir force; then, we try to compute the solution of this anharmonic nonlinear system. Particularly, we try to estimate the stiffness coefficient of this micro spring. An approximate numerical sinusoidal solution is found that shows, for reasonable values of the stiffness, the turning point of the approximately oscillating motion is very near to the initial equilibrium position of the element under consideration; therefore, the vibration of the element is not so troublous for the whole system.

\*Corresponding Author:
Email: razmi@qom.ac.ir
Tel.:+ 98 251 285 4972, Fax: +98 251 285 4972

## *Nonlinear (anharmonic) casimir oscillator*

Consider a simple model consisting of a spring obeying Hook's law connected between two, one is fixed and the other is moving, parallel conducting plates. The spring has an elastic constant $k$ with a free length of $x_0$ (see Fig. 1). The moving plate, in addition to the well-known restoring force of the spring, is under the influence of the Casimir force (pressure) too. In one dimension, just perpendicular to the area of the plate and parallel to the spring length, the unit area of the moving plate is under government of the following equation of motion:

$$\rho_S \frac{d^2 x(t)}{dt^2} + \frac{\gamma}{A}\frac{dx}{dt} + \frac{k}{A}(x - x_0) = \frac{-\pi^2 \hbar c}{240 x^4} \quad (1)$$

where $\rho_S$ is the mass of the unit area ($A$) of the moving plate, and $\gamma$ is the possible damping coefficient of the spring.

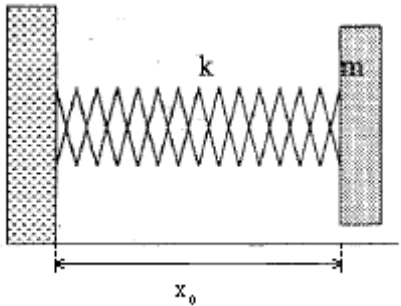

**Fig.1.** A micro spring connected between two parallel conducting plates

Neglecting possible damping ($\gamma = 0$) is a good approximation at the micro (nano) world scales; thus:

$$\ddot{x} + \frac{k}{A \rho_S}(x - x_0) = \frac{-\pi^2 \hbar c}{240 \rho_S x^4} \quad (2)$$

In the limit $\hbar \to 0$, the above equation reduces to the famous linear harmonic oscillator with its well-known solution(s); however, in the presence of the quantum Casimir effect, it is a nonlinear differential equation with no known exact analytical solution.

Since the anharmonic oscillator under consideration is a conservative system, we can write the following energy equation:

$$\frac{1}{2}\rho_S(\frac{dx}{dt})^2 + \frac{1}{2}\frac{k}{A}(x-x_0)^2 - \frac{\pi^2 \hbar c}{720 x^3} = E \quad (3)$$

$$(\frac{dx}{dt})^2 = -\frac{k}{A\rho_S}(x-x_0)^2 + \frac{\pi^2 \hbar c}{360 \rho_S x^3} + \frac{2}{\rho_S}E \quad (4)$$

At the initial time $t = 0$, the total energy $E$ is equal to the Casimir energy ($E_0 = \frac{-\pi^2 \hbar c}{720 x_0^3}$); thus:

$$(\frac{dx}{dt})^2 = \frac{k}{A\rho_S}[-(x-x_0)^2 + \frac{\pi^2 \hbar c}{360 k x^3} - \frac{\pi^2 \hbar c}{360 k x_0^3}] \quad (5)$$

The left-hand side of equation (5) is positive definite; therefore:

$$-(x-x_0)^2 + \frac{\pi^2 \hbar c A}{360 k}(\frac{1}{x^3} - \frac{1}{x_0^3}) \geq 0 \Rightarrow x \leq x_0 \quad (6)$$

In the absence of the Casimir pressure, $x_0$ is the equilibrium center of oscillation; but now, the moving wall, under influence of the Casimir effect, moves towards the fixed wall up to a minimum distance (turning point) $x_{\min}$ and then, because of the spring restoring force, moves backwards. Using (6), it is clear that:

$$x_{\min} < x_0 \quad (7)$$

Because of the nonlinearity of the problem, turning point(s) have no fixed position(s). This is because the problem has not any completely periodic solution. Indeed, the first round of forth and back movement is not repeated in the next (second) round and so on. For the particular situation we consider in the following, the approximate numerical solution is sinusoidal between $x_{\min}$ (very near to $x_0$) and $x_0$. This shows that the variable turning points are collected in a very small interval of positions.

Now, let expand the total potential energy function about its minimum (equilibrium) point $x_{min}$ (knowing $x_{min} < x_0$) up to the second order of $x$:

$$V_{tot}(x) = \frac{1}{2}\frac{k}{A}(x-x_0)^2 - \frac{\pi^2 \hbar c}{720 x^3}$$
$$\approx V(x_{min}) + \frac{1}{2}\left.\frac{d^2 V(x)}{dx^2}\right|_{x_{min}}(x-x_{min})^2 \quad (8)$$

where $x_{min}$, is a solution of $\frac{dV}{dx} = 0$:

$$x_{min}^4(x_0 - x_{min}) = \frac{\pi^2 \hbar c A}{240 k} \quad (9)$$

For the minimum point, in addition to the above relation, the following condition should be also satisfied:

$$\left.\frac{d^2 V(x)}{dx^2}\right|_{x_{min}} > 0 \quad (10)$$

This leads to:

$$k > \frac{\pi^2 \hbar c A}{60 x_{min}^5} \quad (11)$$

From (9) and (11), it is found that:

$$x_{min} > \frac{4}{5}x_0 \quad (12)$$

The inequality (11) (or (12)) is a stability criterion condition. The inequality (12) shows that $x_{min}$ should be near to $x_0$; thus, the expansion of the potential energy converges enough rapidly so that the approximate expansion (8) is well valid.

To find a numerical solution for $x(t)$, we should choose some proper numerical values for the constants of the problem. The value of $x_0$ is considered at the order of 1 micrometer ($x_0 \approx 10^{-6} m$). To be able to apply the simple Casimir force, the area $A$ should be much larger than the second power of separation distance. Assuming $A \geq 100(x_0)^2$, based on (11) and (12), the numerical value of the stiffness constant should satisfy $k \geq 10^{-6}\frac{N}{m}$. We assume $k \sim 1\frac{N}{m}$; this is at the order of currently known greatest possible value [8]. The conducting plates are considered to be made of copper ($\rho_{cu} = 8920\frac{kg}{m^3}$) with a mean thickness of about $1\mu m$; therefore, the mean value of the surface mass density is $\rho_s \sim 8.92 \times 10^{-3}\frac{kg}{m^2}$. Using these data, the nonlinear differential equation (2) takes the following form:

$$\ddot{x} + 1.121 \times 10^{14}(x - 10^{-6}) + \frac{1.459 \times 10^{-25}}{x^4} = 0 \quad (13)$$

Assuming $\dot{x}_0 = 0$, a graphic numerical solution is found as in the Fig. 2.

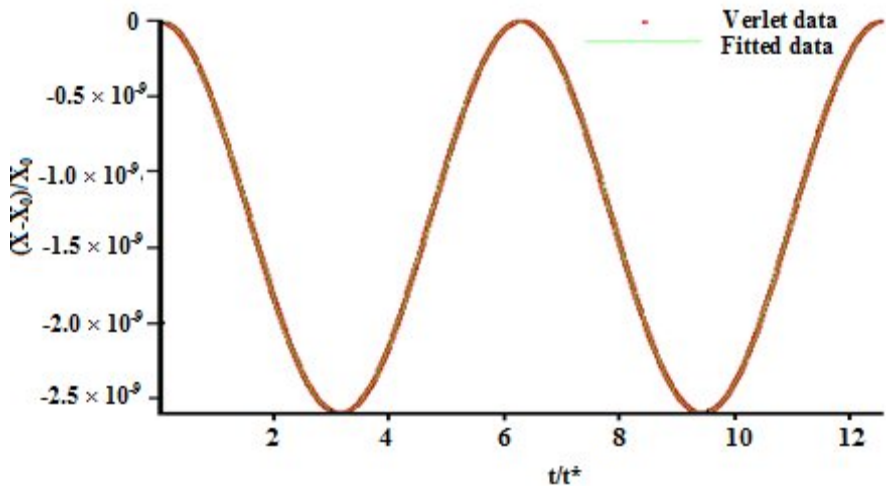

**Fig. 2.** Graphic numerical solution for $x(t)$

This numerical solution has been found using Verlet algorithm for finding the numerical solution of the equation

$$\ddot{x} = a = -b(x - x_0) - \frac{c}{x^4} \quad (14)$$

for which:

$$x(t + \delta t) = 2x(t) - x(t - \delta t) + a\delta t^2 + O(\delta t^4).$$

Working in natural system of units (dimensionless form) of the equation, length unit, $l^*$, is equal to $x_0$ and the time unit, $t^*$, is equal to $b^{-1/2}$. From (13), it is clear that $b = 1.121 \times 10^{14}$ $sec^{-2}$, and $c = 1.459 \times 10^{-25}$. Now, it is found that $x(t)$ vibrates between $x_0$ and $0.9999999974\, x_0$ periodically. In Fig. 2, $(x - x_0)/x_0$ has been plotted versus normalized time $t/t^*$. One can simply fit $x(t)$ with the simple function:

$$x(t)/x_0 = (1 + Amp \times (\cos(\omega t/t^*) - 1)) \quad (15)$$

With $r^2 = 0.999986$. $Amp = (1.302 \pm 0.001) \times 10^{-9}$ and $\omega = 1.000 \pm 0.001$.

## Results and discussion

The nonlinear (anharmonic) Casimir oscillator studied here is intended to be connected between the elements of micromechanical systems which are under the influence of the Casimir stiction. Although there isn't an exact analytical solution for the dynamics of the motion of the micro spring, an approximate numerical solution has been found. The stability criterion condition (11) is satisfied well by data corresponding to currently known world of micromechanical systems.

As it is seen from the stability criterion condition (12) as well as the particular numerical solution found in (15), $x_{min}$ is very near to $x_0$. This shows that the connection of the micro springs between the pieces of micromechanical systems does not disturb these systems' operations. Mathematically, having a $x_{min}$ near to $x_0$ guarantees the convergence of the series expansion (8).

Since the basic idea behind the design proposed here is to prevent the Casimir stiction in micromechanical systems by restoring force of the micro springs, the greatest possible value for the stiffness constant ($k \sim 1\frac{N}{m}$) has been considered not only to satisfy the criterion condition (11) confidentially but also to be at the highest level of managing the problem. Although the value of the stiffness constant really depends on the mass, the geometrical structure of the micro spring, we can assume it is averagely constant (i.e. we can consider an average value for $k$).

The last point, we should mention is about the singularity of the Casimir force/pressure at $x = 0$. Although the Casimir effect at very short distances has a different (e.g. nonretarded effects, van der wals (Lifshitz) considerations) behavior than the famous ideal form, the situation we have studied for the dynamics governing the motion of the moving plate corresponds to the "long range" (retarded) form of the Casimir force. The final solution shows that the moving plate doesn't approach the fixed plate so much (the amplitude of the oscillation is very small).

## Conclusion

Although the idea proposed here is a theoretical plan, it can be used to prevent the Casimir stiction in micromechanical and micro electromechanical systems (MEMS) experimentally. The parameters' values used here are compatible with the current world of nanotechnology and experimentally realistic data.